\documentclass[11pt]{article}

\usepackage[margin=1in]{geometry}
\usepackage{graphicx}
\usepackage{multirow}
\usepackage{amsmath,amssymb,amsfonts}
\usepackage{booktabs}
\usepackage{float}
\usepackage[section]{placeins}
\usepackage{xcolor}
\usepackage[hidelinks]{hyperref}
\usepackage{tikz}
\usetikzlibrary{positioning,arrows.meta,fit,backgrounds}
\usepackage{natbib}
\makeatletter

\providecommand{\bmhead}[1]{\paragraph*{\textbf{#1}}}
\providecommand{\keywords}[1]{\par\noindent\textbf{Keywords:} #1\par}
\makeatother

\title{Investigating Quantum-Embedded Transformers on Classical Datasets for Cross-Modality Classification}

\author{Hao-Yuan Chen\\
\small Computer Science, University of London\\
\small \texttt{hc118@student.london.ac.uk}}

\date{}

\begin{document}

\maketitle

\begin{abstract}
We test whether a parameterized quantum circuit (PQC) improves a hybrid
quantum-classical model's performance on classical datasets, using an
interface-matched classical map as the control while holding all other
components fixed. Our architecture, Quantum-Embedded Attention (QEA), uses a
learnable projector to compress backbone features into an $n_q$-dimensional
angle vector, a shallow PQC to map those angles to one- and two-qubit Pauli
expectations, and a classical attention decoder to produce class logits. We
hypothesized the PQC would improve accuracy or seed-to-seed stability over a
classical map with matched input/output dimensions. We test this with an
interface-matched $2\times2$ factorial on Breast Cancer Wisconsin at
$n_q\in\{4,8\}$, independently swapping the PQC for a classical map and the
attention decoder for a linear head, across five paired seeds per cell. Three
of four paired quantum-minus-classical $95\%$ confidence intervals include
zero; the fourth, a $+1.63$ percentage-point contrast for the attention
decoder at $n_q=4$, reverses sign at $n_q=8$ and does not survive correction
across the four contrasts. The experiment thus shows no consistent PQC
contribution and cannot establish equivalence. A five-dataset cross-modality
grid shows comparable accuracy on AG~News, Breast Cancer Wisconsin, and
BirdCLEF but a large deficit on CIFAR-10; these cells are not
interface-matched and are interpreted descriptively. We report all planned
canonical runs, distinguish current Pauli-readout results from legacy
probability-readout experiments, and analyze bottleneck, simulation,
finite-shot, and noise limitations. The results do not establish a quantum
advantage; they demonstrate why controlled component attribution is necessary
before crediting a hybrid model's performance to its quantum layer.

\end{abstract}


\section{Introduction}
\label{sec:intro}

Hybrid quantum--classical models are a practical way to study quantum machine
learning (QML) on present hardware while retaining classical feature extractors
and optimizers~\cite{preskill2018nisq,benedetti2019pqc}. Restricting the quantum
component to a small parameterized circuit (PQC) after a classical projector
avoids loading the raw high-dimensional input as amplitudes. Keeping the circuit
shallow also reduces---but does not eliminate---the trainability risks associated
with deep PQCs, including barren plateaus~\cite{mcclean2018bp}.

\paragraph{Motivation and hypothesis.}
Two properties make a PQC an appealing module to test inside a modern network.
First, its input interface can be modality-agnostic: after a classical backbone
has mapped a spectrogram, image, tabular vector or sentence to a feature vector,
the same fixed-width circuit design can act on the projected features. Second, a
PQC supplies a trainable non-linear map whose parameter count can grow slowly
with circuit width. Neither property implies that the map is more useful than a
classical alternative, and the exponential Hilbert-space dimension alone does not
establish useful capacity or computational advantage. The scientific question is
therefore comparative. Our guiding hypothesis is \emph{(H): after the same learned
projector, a PQC improves accuracy or seed-to-seed stability relative to a
classical map with the same input and output dimensions.} Its empirical prediction
is a non-zero paired performance difference when only that module is exchanged.

\paragraph{Where to insert the circuit.}
The interface between a backbone and its classifier is a natural test location:
the input has already been compressed, and the inserted module can be exchanged
without changing the data pipeline. We call the design pattern a \emph{quantum
embedding layer} and the architecture studied here \emph{Quantum-Embedded
Attention (QEA)}. An attention decoder acts on the measured channels after the
circuit; it does not implement quantum attention. This work extends a preliminary
single-qubit, single-dataset study by the same authors~\cite{chen2024qet}
(Section~\ref{sec:related}) with a multi-qubit Pauli readout, a broader modality
study, and a controlled intervention on the quantum layer. The placement is thus
motivated by testability as much as by hardware constraints: it exposes a single
quantum module whose classical surrogate can occupy the same interface.

The contributions of this paper are:
\begin{enumerate}
\item \textbf{A self-contained QEA specification}: backbone $\to$ learnable
  projector $\to$ data-dependent PQC $\to$ one- and two-body Pauli readout $\to$
  classical attention decoder. The qubit count $n_q$ is decoupled from raw input
  dimensionality, and every stage is defined mathematically
  (Section~\ref{sec:method}).
\item \textbf{An interface-matched factorial that tests the quantum
  contribution.} At fixed $n_q$ we independently replace the PQC with a classical
  map of identical input/output dimensions and replace the attention decoder with
  a linear head. The projector, readout interface, seeds and training budget are
  held fixed. We report paired effect intervals and do not infer equivalence from
  overlapping marginal intervals.
\item \textbf{A five-dataset cross-modality evaluation and complete accounting
  of the controlled runs.} QEA shows no consistent advantage over classical
  controls and fails strongly on CIFAR-10. We report the incomplete chemistry
  cell, unsuccessful runs, scaling costs and the distinction between legacy
  probability readout and the current Pauli-readout protocol instead of filling
  missing evidence by extrapolation.
\end{enumerate}

\section{Related work}
\label{sec:related}

\paragraph{Variational encodings.}
Variational quantum classifiers use data-dependent circuits and trainable gates to
fit decision boundaries~\cite{schuld2020vqc,havlicek2019supervised}. Their
function class depends strongly on encoding. Data re-uploading interleaves repeated
encodings with trainable blocks~\cite{perez2020data}, and the resulting models can
be analyzed as truncated Fourier series whose accessible frequencies depend on
the encoding~\cite{schuld2021effect}. P\'erez-Salinas et
al.~\cite{perez2020data} compare the repeated processing of a single-qubit
classifier with a one-hidden-layer neural network, in which hidden units likewise
receive repeated copies of the input. Moreover, a single qubit is efficiently
classically simulable. Strong single-qubit performance is therefore insufficient
evidence of a quantum contribution and motivates the explicit surrogate used in
Section~\ref{sec:methods}.

\paragraph{Relation to the authors' prior work.}
The closest predecessor is a study by the same authors~\cite{chen2024qet} that
inserted a single-qubit circuit after a vision transformer and evaluated one
BirdCLEF task, reporting an approximately three-point median-$F_1$ improvement.
That comparison changed the hybrid pipeline as a whole and did not isolate the
projector, circuit and decoder. The present work differs by using multi-qubit
Pauli readout, a classical attention decoder, five current-protocol datasets and,
most importantly, an interface-matched $2\times2$ factorial. The earlier paper
asked whether a hybrid model can work; this paper asks whether a measured change
persists when the quantum module alone is exchanged. The current evidence does
not show a consistent effect across the tested BCW settings.

\paragraph{Quantum kernels---and why this is not one.}
Quantum kernel methods embed inputs into quantum states and evaluate pairwise
similarities before fitting a classical kernel
learner~\cite{havlicek2019supervised,schuld2021kernel}. A common example is the
fidelity
\begin{equation}
k(x_i,x_j)=|\langle\phi(x_i)|\phi(x_j)\rangle|^2.
\end{equation}
QEA computes no inter-example similarity or Gram matrix. Each input is mapped to a vector of
local Pauli expectations and decoded independently. This distinction also locates the work relative to current evidence. Schnabel and
Roth's benchmark of more than $20{,}000$ fidelity and projected quantum-kernel
models across 64 datasets found strong sensitivity to encoding, kernel and
hyperparameter choices and no universal quantum advantage~\cite{schnabel2025scrutiny}.
Our model is not one of those kernels, but their emphasis on classical controls
and component-level analysis applies directly.

\paragraph{Quantum transformers and hybrid feature extractors.}
The recent quantum-transformer survey distinguishes PQC-based hybrid designs from
fault-tolerant quantum-linear-algebra proposals and highlights scalability,
benchmarking and trainability gaps~\cite{zhang2025qtsurvey}. Quantum vision
transformers place PQCs inside or around attention operations~\cite{cherrat2024qsa},
whereas quanvolutional networks use small circuits as local feature
extractors~\cite{henderson2020quanvolutional}. QEA is simpler: its attention is
fully classical, and the PQC occupies a post-backbone embedding interface. This
placement is chosen because the circuit can be exchanged without redefining the
task, not because it is expected to accelerate classical self-attention.

Frameworks such as Qiskit Machine Learning~\cite{qiskitml} and
PennyLane~\cite{bergholm2018pennylane} support differentiation through quantum
layers. The reported experiments use the batched PyTorch statevector
implementation specified in Section~\ref{sec:qel}; framework selection is not a
novelty claim.

\section{Method}
\label{sec:method}

\subsection{Architecture overview}

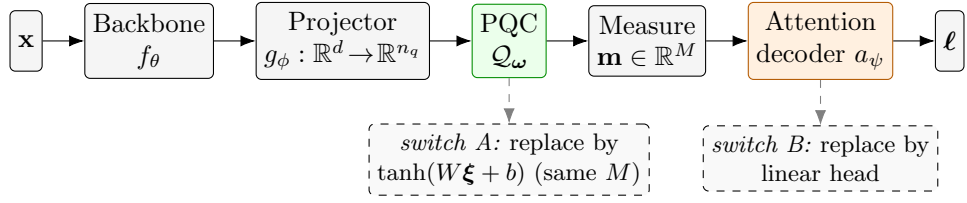
\begin{figure}[t]
\centering
\begin{tikzpicture}[
  font=\small,
  >={Latex[length=2mm]},
  box/.style={rounded corners=2pt,draw,minimum height=8mm,inner xsep=3pt,align=center},
  cl/.style={box,fill=black!4},
  qu/.style={box,fill=green!8,draw=green!55!black},
  atn/.style={box,fill=orange!12,draw=orange!70!black},
  alt/.style={box,dashed,fill=gray!5,align=center,font=\footnotesize},
  node distance=5.5mm,
]
\node[cl] (x) {$\mathbf{x}$};
\node[cl,right=of x] (bb) {Backbone\\$f_\theta$};
\node[cl,right=of bb] (pr) {Projector\\$g_\phi:\mathbb{R}^d\!\to\!\mathbb{R}^{n_q}$};
\node[qu,right=of pr] (q) {PQC\\$\mathcal{Q}_{\boldsymbol{\omega}}$};
\node[cl,right=of q] (m) {Measure\\$\mathbf{m}\in\mathbb{R}^{M}$};
\node[atn,right=of m] (a) {Attention\\decoder $a_\psi$};
\node[cl,right=of a] (l) {$\boldsymbol{\ell}$};
\draw[->] (x)--(bb); \draw[->] (bb)--(pr); \draw[->] (pr)--(q);
\draw[->] (q)--(m); \draw[->] (m)--(a); \draw[->] (a)--(l);
\node[alt,below=6mm of q] (qc) {\emph{switch A:} replace by\\$\tanh(W\boldsymbol{\xi}+b)$ (same $M$)};
\node[alt,below=6mm of a] (ac) {\emph{switch B:} replace by\\linear head};
\draw[->,dashed,gray] (q)--(qc);
\draw[->,dashed,gray] (a)--(ac);
\end{tikzpicture}
\caption{The core QEA pipeline of Eq.~\eqref{eq:qet}. A classical backbone and
projector reduce the input to an $n_q$-dimensional angle vector, a
parameterized quantum circuit produces a vector of measurement statistics, and
an attention decoder maps those statistics to class logits. The two dashed
switches define the interface-matched factorial of
Section~\ref{sec:experiments}: switch~A replaces the circuit with an equal-width
classical map, and switch~B replaces the attention decoder with a linear head.
The broader exploratory grid uses the QEA-R variant, which additionally
concatenates the projected angles as a residual token stream; that variant is
reported separately and is not used for component attribution.}
\label{fig:method}
\end{figure}

QEA maps a raw input $\mathbf{x}$ to class logits $\boldsymbol{\ell}$ through four
composed stages,
\begin{equation}
\mathbf{x}
\;\xrightarrow{\;f_\theta\;}\; \mathbf{h}\in\mathbb{R}^{d}
\;\xrightarrow{\;g_\phi\;}\; \boldsymbol{\xi}\in\mathbb{R}^{n_q}
\;\xrightarrow{\;\mathcal{Q}_{\boldsymbol{\omega}}\;}\; \mathbf{m}\in\mathbb{R}^{M}
\;\xrightarrow{\;a_\psi\;}\; \boldsymbol{\ell}\in\mathbb{R}^{C},
\label{eq:qet}
\end{equation}
where $f_\theta$ is a modality-specific classical backbone, $g_\phi$ a learnable
projector, $\mathcal{Q}_{\boldsymbol{\omega}}$ a quantum embedding that returns a
vector $\mathbf{m}$ of measurement statistics, and $a_\psi$ an attention decoder.
Figure~\ref{fig:method} shows the core pipeline together with the two ablation
switches used in Section~\ref{sec:experiments}. All parameters
$\{\theta,\phi,\boldsymbol{\omega},\psi\}$ are trained jointly by minimizing the
cross-entropy loss with Adam~\cite{kingma2015adam}; gradients through
$\mathcal{Q}_{\boldsymbol{\omega}}$ are analytic (Section~\ref{sec:qel}).

\subsection{Classical backbone and projector}

The backbone $f_\theta$ is chosen per modality: ResNeXt-50~\cite{xie2017resnext}
for audio spectrograms and images, a multilayer perceptron for tabular,
chemistry and physics feature vectors, and a frozen
MiniLM~\cite{wang2020minilm} sentence encoder followed by a small MLP for text.
It outputs a feature $\mathbf{h}=f_\theta(\mathbf{x})\in\mathbb{R}^d$. The
projector is a two-layer network
\begin{equation}
g_\phi(\mathbf{h}) = W_2\,\sigma\!\left(W_1\mathbf{h}+b_1\right)+b_2,
\qquad \sigma=\mathrm{GELU},
\end{equation}
with $W_2\in\mathbb{R}^{n_q\times h}$, that compresses $\mathbf{h}$ to one angle
per qubit. Crucially, $n_q$ is a design choice \emph{independent of $d$}: a
$224\times224$ image ($d$ from the backbone) and a $30$-dimensional tabular
vector both terminate in the same $n_q$-wide bottleneck. The angles are bounded
before encoding, $\boldsymbol{\xi}=\pi\tanh\!\big(s\odot g_\phi(\mathbf{h})\big)$,
with an optional learnable per-qubit scale $s$. The controlled factorial fixes
$s=\mathbf{1}$; the exploratory QEA-R grid learns $s$.

\subsection{Quantum embedding layer}
\label{sec:qel}

The quantum stage applies a data-dependent unitary to the $n_q$-qubit ground
state and reads out a fixed set of expectation values. The unitary factorizes
into a data-encoding feature map $\mathcal{F}$ and a trainable ansatz
$\mathcal{A}$,
\begin{equation}
U(\boldsymbol{\xi};\boldsymbol{\omega})
  = \mathcal{A}(\boldsymbol{\omega})\,\mathcal{F}(\boldsymbol{\xi}),
\qquad
|\psi(\boldsymbol{\xi};\boldsymbol{\omega})\rangle
  = U(\boldsymbol{\xi};\boldsymbol{\omega})\,|0\rangle^{\otimes n_q}.
\end{equation}
The feature map is a layer of single-qubit rotations optionally followed by
entangling $ZZ$ rotations,
\begin{equation}
\mathcal{F}(\boldsymbol{\xi})
 = \prod_{r=1}^{R}\Bigg[
   \prod_{(i,j)\in E} e^{-i\,\xi_i\xi_j\,Z_iZ_j}
   \prod_{i=1}^{n_q} e^{-i\,\xi_i\,Z_i}
   \prod_{i=1}^{n_q} H_i \Bigg],
\end{equation}
where $R$ is the number of encoding repetitions and $E$ the entangling graph;
setting $E=\varnothing$ recovers a product ($Z$) feature map. Repeating the same
angles within the feature map is data re-uploading in the broad sense, and $R$
changes the accessible Fourier frequencies~\cite{perez2020data,schuld2021effect}.
The controlled factorial uses $R{=}2$ followed by the ansatz. The exploratory
QEA-R grid uses the same two-repetition feature-map block before each trainable
ansatz layer. We therefore do not describe either protocol as ``single
encoding''; the only single interface is the location at which the classical
projector supplies $\boldsymbol{\xi}$ to the circuit. The ansatz is a
hardware-efficient block of parameterized $Y$-rotations and nearest-neighbour
$\mathrm{CNOT}$s,
\begin{equation}
\mathcal{A}(\boldsymbol{\omega})
 = \prod_{r=1}^{L}\Bigg[
   \prod_{i=1}^{n_q-1}\mathrm{CNOT}_{i,i+1}
   \prod_{i=1}^{n_q} R_Y(\omega_{r,i})\Bigg]
   \prod_{i=1}^{n_q} R_Y(\omega_{0,i}),
\end{equation}
with depth $L$ (the \texttt{RealAmplitudes} family; \texttt{EfficientSU2}, used
in one ablation, adds $R_Z$ rotations). The number of trainable circuit
parameters is $n_q(L{+}1)$, i.e.\ \emph{linear} in the qubit count while the
state space is $2^{n_q}$-dimensional.

\paragraph{Readout.}
Rather than returning the full $2^{n_q}$-dimensional probability vector, whose
dimension grows exponentially and which is expensive to estimate on hardware, we
read out a polynomial-size vector of Pauli expectation values,
\begin{equation}
\mathbf{m} = \big(\langle Z_i\rangle_{i}\,,\ \langle Z_iZ_j\rangle_{(i,j)\in E}\big),
\qquad M = n_q + |E|,
\label{eq:readout}
\end{equation}
so $\mathbf{m}\in\mathbb{R}^{M}$ with $M=\mathcal{O}(n_q^2)$. This avoids an
exponential classical decoder but does not, by itself, establish a quantum
speedup; Section~\ref{sec:discussion} separates readout, simulation and hardware
costs. All reported expectation values are computed by a batched, exact
statevector simulator, and gradients are obtained by automatic differentiation
through that simulator. Hardware execution would require finite-shot expectation
estimates and a hardware-compatible gradient estimator such as parameter shift;
it is not evaluated here.

\subsection{Attention decoder}
\label{sec:head}

The decoder $a_\psi$ treats the readout as a short sequence of tokens. Each
measurement channel $m_k$ is lifted to a token
$t_k = e_k + r(m_k)\in\mathbb{R}^{D_a}$, where $e_k$ is a learned per-channel
embedding and $r$ a linear projection of the scalar $m_k$. A stack of standard
multi-head self-attention and feed-forward blocks transforms the tokens
$\{t_k\}$, and the sequence is pooled and passed to a small MLP classifier:
\begin{equation}
\mathbf{z} = \mathrm{Pool}\big(\mathrm{Attn}_\psi(\{t_k\}_{k=1}^{M})\big),
\qquad
\boldsymbol{\ell} = \mathrm{MLP}_\psi(\mathbf{z}).
\end{equation}
The decoder is entirely classical; its role is to model interactions among
measurement channels that a single linear layer cannot. Its contribution is
isolated by switch~B in Figure~\ref{fig:method}, which replaces
$a_\psi$ with a single linear map $\boldsymbol{\ell}=W\mathbf{m}+b$.

\paragraph{Exploratory residual variant.}
The cross-modality grid predates the controlled factorial and uses a stronger
variant denoted QEA-R. After an identity-initialized learned mixer of the Pauli
channels, QEA-R concatenates the projected angle vector $\boldsymbol{\xi}$ to
$\mathbf{m}$ before attention. This classical residual path can bypass the PQC;
accordingly, QEA-R is useful as a descriptive hybrid baseline but cannot attribute
its predictions to the circuit. All component-attribution statements in this
paper come from the core, residual-free factorial in
Section~\ref{sec:methods}.

\subsection{Qubit count and circuit depth}
\label{sec:nq}

Because $n_q$ is decoupled from $d$, it is a free bottleneck that we sweep as an
ablation rather than fix to the input size. Following McClean et
al.~\cite{mcclean2018bp} we keep the ansatz shallow ($L\in\{1,2\}$) to limit
barren-plateau risk; the consequences of this choice for expressivity are
discussed in Section~\ref{sec:discussion}. To keep the comparison against the
Hybrid baseline free of a qubit-count confound, both methods are evaluated at
\emph{identical} $n_q$ in the interface-matched study
(Section~\ref{sec:experiments}).

\section{Experimental setup}
\label{sec:experiments}

The evaluation separates two questions. The \emph{interface-matched factorial}
tests hypothesis (H) by changing one architectural component at a time; it is the
only analysis used for component attribution. The \emph{cross-modality grid}
asks whether an exploratory residual variant can be trained on heterogeneous
inputs. Because its historical baselines differ in readout width and training
details, that grid is descriptive rather than causal.

\subsection{Datasets and analysis status}
\label{sec:datasets}

The study repository contains loaders for six source datasets
(Table~\ref{tab:datasets}). Current Pauli-readout results are complete for five:
BirdCLEF-2021 ``nocall'' audio detection~\cite{birdclef2021}, CIFAR-10
images~\cite{krizhevsky2009cifar}, Breast Cancer Wisconsin (Diagnostic,
BCW)~\cite{wolberg1995bcw}, SUSY high-energy physics~\cite{baldi2014susy}, and
AG~News text classification~\cite{zhang2015agnews}. The available QM9
chemistry rows~\cite{ramakrishnan2014qm9} use an older two-qubit,
probability-readout implementation and are not pooled with the current protocol.
This revision therefore reports a five-dataset grid, not a six-dataset result.
Appendix~\ref{app:data} gives the actual sample counts and split construction.

\begin{table}[t]
\centering
\small
\caption{Dataset scope. ``Current'' denotes inclusion in the five-dataset
Pauli-readout grid; QM9 is retained as a documented source dataset but excluded
from current aggregate claims because only legacy-protocol rows are available.}
\label{tab:datasets}
\begin{tabular}{lllll}
\toprule
Modality & Dataset & Task & Backbone & Analysis status \\
\midrule
Audio     & BirdCLEF~\cite{birdclef2021}        & binary   & ResNeXt-50   & Current \\
Image     & CIFAR-10~\cite{krizhevsky2009cifar} & 10-class & ResNeXt-50   & Current \\
Tabular   & BCW~\cite{wolberg1995bcw}           & binary   & MLP          & Current + factorial \\
Chemistry & QM9~\cite{ramakrishnan2014qm9}      & 5-bin    & MLP          & Legacy protocol only \\
HEP       & SUSY~\cite{baldi2014susy}           & binary   & MLP          & Current \\
NLP       & AG~News~\cite{zhang2015agnews}      & 4-class  & MiniLM + MLP & Current \\
\bottomrule
\end{tabular}
\end{table}

\subsection{Interface-matched factorial}
\label{sec:methods}

We define every controlled model by the four stages of Eq.~\eqref{eq:qet}:
backbone $f_\theta$, projector $g_\phi$, embedding
$\mathcal{Q}_{\boldsymbol{\omega}}$, and decoder $a_\psi$. Two switches act on
the last two stages:
\begin{description}
\item[Switch A (embedding).] The PQC is replaced by
  $\widetilde{\mathcal{Q}}(\boldsymbol{\xi})=
  \tanh(W\boldsymbol{\xi}+b)$, which has the same $n_q$-dimensional input and
  $M$-dimensional output as Eq.~\eqref{eq:readout}. The two maps do not have
  identical parameter counts, so we call the control \emph{interface-matched},
  not parameter-matched; total trainable counts are reported in
  Appendix~\ref{app:arch}.
\item[Switch B (decoder).] The attention decoder of
  Section~\ref{sec:head} is replaced by a single linear classifier on the same
  readout vector.
\end{description}
The four combinations are QEA (PQC + attention), Hybrid QNN (PQC + linear),
Classical + Attention, and Classical + Linear. At each
$n_q\in\{4,8\}$, all four use the same BCW split for a given seed, the same MLP
backbone, projector, $M=n_q(n_q+1)/2$ readout interface, optimizer schedule and
40-epoch budget. Five seeds ($42,123,31,2024,7$) are paired across cells. Thus a
within-seed row comparison localizes the architectural intervention to switch~A;
it does not turn a non-significant result into proof of equivalence.

\subsection{Exploratory cross-modality grid}
\label{sec:cross-grid}

The broader grid reports four historical model families: a plain backbone + MLP
classifier (Classical), a projector + classical readout + attention model
(Classical + Attention), a two-qubit PQC + linear head (Hybrid QNN), and the
eight-qubit residual quantum model QEA-R defined in Section~\ref{sec:head}. These
columns are intentionally not presented as an ablation: the Hybrid width is two
qubits, the classical-attention readout has $2^{8}$ channels, and QEA-R has 36
Pauli channels plus an eight-angle residual. The table answers whether each
configured pipeline trains, not which component caused a difference.

\subsection{Training, uncertainty and run accounting}
\label{sec:protocol}

All models use Adam~\cite{kingma2015adam} with nominal learning rate $10^{-3}$,
weight decay $10^{-6}$, gradient clipping at $5$, a cosine warm-restart schedule,
and selection of the best validation checkpoint before one held-out test
evaluation. The factorial uses 40 epochs for every cell. Cross-grid CSVs record
30 epochs for most completed runs; four historical rows ended at 15--25 epochs
(one AG~News Hybrid, two BirdCLEF Classical + Attention, and one CIFAR-10
QEA-R). They are retained only in the descriptive grid and are not evidence for
component attribution. Marginal table intervals are two-sided $95\%$ Student-$t$
intervals over seeds. For the factorial, the primary uncertainty is the
Student-$t$ interval of the paired quantum-minus-classical differences. With only
five pairs, these intervals are imprecise; no equivalence margin was prespecified.

A canonical run is marked \emph{collapsed} when validation accuracy equals the
majority-class rate at three consecutive scheduled evaluations. Smoke tests are
diagnostics rather than planned analysis units. A non-finite loss raises an error
and produces no result row; the earlier SUSY errors were traced to a truncated
non-finite input row, after which all canonical SUSY Classical + Attention cells
were rerun with a loader that drops non-finite rows before splitting. Run counts,
duplicate-resolution rules and incomplete budgets are reported in
Section~\ref{sec:accounting}; no failed canonical run is silently replaced by a
different seed.

\subsection{Reproducibility}
\label{sec:repro}

Each canonical cell is specified by a versioned configuration file, and the
run-level CSVs, dependency lockfile and scripts that regenerate the tables and
figures accompany the revision. Operational commands and legacy-code notes remain
in the repository documentation rather than the scientific narrative.

\section{Results}
\label{sec:results}

We report the controlled factorial first, followed by the run ledger and the
exploratory cross-modality grid. This order reflects evidentiary weight: only the
factorial changes one interface at a time.

\subsection{Paired component attribution on BCW}
\label{sec:isolation}

Table~\ref{tab:isolation} reports held-out accuracy for the four combinations of
embedding and decoder at $n_q\in\{4,8\}$. Marginal means are accompanied by
$95\%$ Student-$t$ intervals; the final column is the within-seed
quantum-minus-classical difference, which is the relevant estimate for switch~A.

\begin{table}[!htbp]
\centering
\caption{Interface-matched factorial on BCW (percentage-point test accuracy,
mean $\pm$ $95\%$ Student-$t$ CI, five paired seeds). The paired effect is
quantum minus classical at fixed decoder and $n_q$. Three intervals include zero.
The $n_q=4$ attention contrast is positive before multiplicity correction, but
the direction does not replicate at $n_q=8$.}
\label{tab:isolation}
\begin{tabular}{llccc}
\toprule
$n_q$ & Decoder & Quantum (PQC) & Classical surrogate & Paired $\Delta_{Q-C}$ \\
\midrule
4 & Attention & 97.21\,$\pm$\,2.62 & 95.58\,$\pm$\,2.58 & +1.63\,$\pm$\,1.29 \\
4 & Linear & 95.81\,$\pm$\,4.16 & 96.05\,$\pm$\,2.62 & -0.23\,$\pm$\,2.14 \\
\midrule
8 & Attention & 95.58\,$\pm$\,5.53 & 96.28\,$\pm$\,1.88 & -0.70\,$\pm$\,5.06 \\
8 & Linear & 97.21\,$\pm$\,1.65 & 96.74\,$\pm$\,1.58 & +0.47\,$\pm$\,0.79 \\
\bottomrule
\end{tabular}

\end{table}

\begin{figure}[!htbp]
\centering
\includegraphics[width=0.92\linewidth]{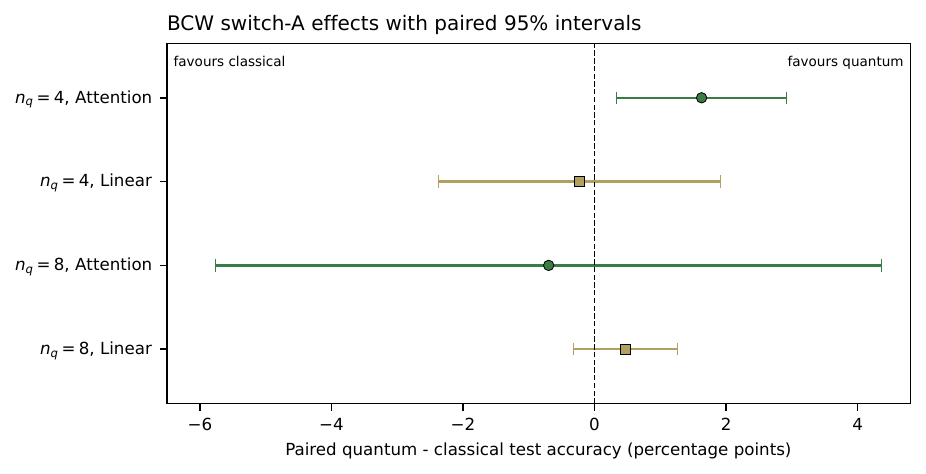}
\caption{Paired quantum-minus-classical switch-A effects in the BCW factorial.
Points denote the mean within-seed difference and error bars the $95\%$
Student-$t$ interval over five paired seeds. The vertical line marks zero. The
plot makes visible both the isolated positive $n_q=4$ attention contrast and its
failure to replicate at $n_q=8$; numerical values are in
Table~\ref{tab:isolation}.}
\label{fig:isolation}
\end{figure}
\FloatBarrier

At $n_q=4$ with attention, QEA exceeds its classical surrogate by $1.63$ points
(paired $95\%$ CI $[0.34,2.92]$, unadjusted paired $t$ test $p=0.025$). The other
three switch-A effects are $-0.23$ points with a linear decoder at $n_q=4$
($[-2.37,1.91]$), $-0.70$ points with attention at $n_q=8$
($[-5.76,4.37]$), and $+0.47$ points with a linear decoder at $n_q=8$
($[-0.33,1.26]$). Applying Benjamini--Hochberg correction to these four
embedding contrasts gives $p_{\mathrm{BH}}=0.100$ for the smallest $p$ value.
The positive $n_q=4$ contrast is therefore a result to replicate, not a stable
quantum effect: it is small relative to the task's $\approx96\%$ ceiling, changes
direction at $n_q=8$, and was selected from four contrasts with only five pairs
each. Conversely, the data are too imprecise for an equivalence claim because no
smallest effect of interest was prespecified.

The decoder contrasts are also inconsistent across widths. On this saturated
dataset, the factorial neither establishes an attention benefit nor supports the
stronger claim that the PQC is interchangeable with a classical map on all tasks.
Its defensible conclusion is narrower: the prespecified architecture did not
produce a consistent switch-A effect across the two tested widths on BCW.

\subsection{Run accounting}
\label{sec:accounting}

All 40 planned factorial runs (four cells $\times$ two widths $\times$ five
seeds) completed without a majority-class collapse. Some CSVs contain repeated
executions of the same seed at shorter budgets; aggregation deterministically
selects the row with the largest recorded epoch count, so duplicates are not
treated as independent observations (Table~\ref{tab:accounting}).

\begin{table}[!htbp]
\centering
\caption{Canonical factorial run accounting after resolving repeated rows by
maximum recorded epoch count. ``Coll.'' is the number meeting the fixed collapse
rule. The all-run and clean-run means coincide because no factorial run
collapsed.}
\label{tab:accounting}
\begin{tabular}{lccccc}
\toprule
Cell & $n_q$ & Planned & Coll. & Acc.\ (all) & Acc.\ (clean) \\
\midrule
Quantum+Attn (QEA) & 4 & 5 & 0 & 97.21 & 97.21 \\
Classical+Attn & 4 & 5 & 0 & 95.58 & 95.58 \\
Quantum+Linear (Hybrid) & 4 & 5 & 0 & 95.81 & 95.81 \\
Classical+Linear & 4 & 5 & 0 & 96.05 & 96.05 \\
\midrule
Quantum+Attn (QEA) & 8 & 5 & 0 & 95.58 & 95.58 \\
Classical+Attn & 8 & 5 & 0 & 96.28 & 96.28 \\
Quantum+Linear (Hybrid) & 8 & 5 & 0 & 97.21 & 97.21 \\
Classical+Linear & 8 & 5 & 0 & 96.74 & 96.74 \\
\bottomrule
\end{tabular}

\end{table}
\FloatBarrier

The five-dataset grid contains 84 planned canonical configurations: 21 each for
Classical, Classical + Attention, Hybrid QNN and QEA-R, reflecting three seeds
for image/audio and five for the other datasets. Classical, Classical + Attention
after the corrected SUSY reload, and QEA-R have $0/21$ collapsed configurations.
Hybrid QNN has $3/21$: one BirdCLEF and two SUSY seeds. These three remain in the
run ledger and are excluded from clean-cell means by the stated rule, leaving two
BirdCLEF and three SUSY Hybrid estimates. Four additional clean rows have shorter
historical budgets, as disclosed in Section~\ref{sec:protocol}. Smoke diagnostics
and repeated infrastructure attempts are not additional seed-level observations.
QM9 is outside this 84-run current-protocol ledger because its available rows use
the legacy probability readout.

\subsection{Exploratory cross-modality grid}
\label{sec:main-results}

\begin{figure}[H]
\centering
\includegraphics[width=\linewidth]{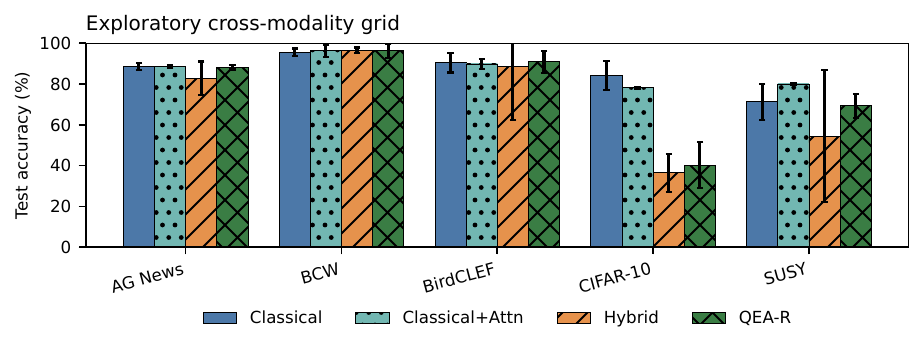}
\caption{Exploratory cross-dataset comparison corresponding to
Table~\ref{tab:main-results}. Error bars are $95\%$ Student-$t$ intervals; fill
patterns distinguish methods without relying on color. The figure shows similar
point estimates on three datasets, a QEA-R deficit on SUSY, and a large failure of
both quantum models on CIFAR-10.}
\label{fig:main-results}
\end{figure}

Table~\ref{tab:main-results} and Figure~\ref{fig:main-results} summarize the five
current datasets. Because the columns are not interface-matched, the comparisons
are descriptive. On AG~News, BCW and BirdCLEF, QEA-R's point estimate is close to
the classical estimates. This cannot be credited to the PQC: QEA-R includes an
angle residual that bypasses it, and the classical-attention column uses a
different readout width.

The two harder datasets argue against a general performance benefit. On
CIFAR-10, QEA-R reaches $40.28\%$ versus $84.11\%$ for the plain Classical model
and $78.14\%$ for Classical + Attention. On SUSY, Classical + Attention reaches
$80.06\%$, compared with $69.35\%$ for QEA-R, $71.29\%$ for Classical, and
$54.30\%$ for the collapse-prone Hybrid QNN. Thus the broad grid contains no
consistent empirical advantage for the quantum path; its clearest result is the
large failure on CIFAR-10.

\begin{table}[!htbp]
\centering
\caption{Exploratory held-out test accuracy (mean $\pm$ $95\%$ Student-$t$ CI).
The columns are historical model families, not a component ablation. AG~News,
BCW and SUSY use five planned seeds; BirdCLEF and CIFAR-10 use three. After the
fixed collapse rule, BirdCLEF and SUSY Hybrid retain two and three seeds,
respectively. QEA-R contains a classical angle residual; the controlled core QEA
comparison is Table~\ref{tab:isolation}.}
\label{tab:main-results}
\begin{tabular}{llcccc}
\toprule
Dataset & Backbone & Classical & Class.+Attn & Hybrid $n_q{=}2$ & QEA-R $n_q{=}8$ \\
\midrule
AG News & MiniLM+MLP & 88.47\,$\pm$\,1.74 & 88.60\,$\pm$\,0.82 & 82.98\,$\pm$\,8.12 & 88.26\,$\pm$\,1.09 \\
BCW & MLP & 95.58\,$\pm$\,1.88 & 96.28\,$\pm$\,3.13 & 96.51\,$\pm$\,1.44 & 96.28\,$\pm$\,3.29 \\
BirdCLEF & ResNeXt-50 & 90.42\,$\pm$\,4.74 & 89.86\,$\pm$\,2.61 & 88.75\,$\pm$\,26.47 & 90.97\,$\pm$\,5.31 \\
CIFAR-10 & ResNeXt-50 & 84.11\,$\pm$\,7.23 & 78.14\,$\pm$\,0.44 & 36.46\,$\pm$\,9.39 & 40.28\,$\pm$\,11.20 \\
SUSY & MLP & 71.29\,$\pm$\,8.89 & 80.06\,$\pm$\,0.38 & 54.30\,$\pm$\,32.39 & 69.35\,$\pm$\,5.84 \\
\bottomrule
\end{tabular}

\end{table}

\FloatBarrier

\subsection{What the seed failures do and do not show}
\label{sec:reliability}

The $3/21$ Hybrid collapse rate is an observed property of this implementation
and schedule, not evidence that linear quantum heads are intrinsically unstable.
The non-collapsed attention models are associated with lower variance on AG~News
and SUSY, but this association is confounded by decoder capacity, readout width
and, for QEA-R, the residual bypass. No controlled hard-task factorial is
available. We therefore report collapse frequency as a reproducibility outcome
and do not label QEA-R ``quantum-robust.''

\section{Discussion}
\label{sec:discussion}

\subsection{What was learned about hypothesis (H)?}
\label{sec:attribution}

Hypothesis (H) predicted that exchanging the classical surrogate for the PQC at
the same interface would improve accuracy or stability. The BCW factorial does
not show a consistent effect across widths or decoders. One of four paired
contrasts is positive before multiplicity correction; the effect changes sign at
the other width and the adjusted test is not significant. This pattern is
compatible with a small task-specific effect, optimization noise, or a false
positive. It is not compatible with a broad claim that the quantum layer is the
source of QEA's performance. It also does not prove that the maps are equivalent:
five seeds on one saturated dataset provide little power to exclude small effects.

A hybrid model performing well is evidence about the full hybrid pipeline, not its
PQC. The $n_q=1$ results make the same point from another direction. A
single-qubit re-uploading classifier is classically simulable, and its processing
can be compared with a one-hidden-layer neural network whose repeated uses of the
input play a role analogous to re-uploads~\cite{perez2020data}. High accuracy at
$n_q=1$ is therefore a useful debugging result but not evidence of quantum
advantage.

\subsection{Attention and the compression bottleneck}
\label{sec:bottleneck}

The current evidence cannot fully separate compression from the subsequent map. QEA-R and Classical + Attention both use an
eight-dimensional projector, but their readouts differ: QEA-R supplies 36 Pauli
channels plus an eight-angle residual, whereas the historical classical control
supplies 256 softmax channels. The gap from plain Classical ($84.11\%$) to
Classical + Attention ($78.14\%$) is consistent with a cost from the bottleneck,
but it also changes the decoder and parameterization. The further drop to QEA-R
($40.28\%$) is likewise not an isolated circuit effect. A decisive image
experiment would use switch~A with identical 36-channel readouts, residual paths,
seeds and budgets on both sides.

The same qualification applies to the role of attention. On BCW, switch~B has no
consistent effect because all four cells are near the task ceiling. On SUSY, the
historical Classical + Attention column is strongest, while on AG~News and
BirdCLEF its point estimate is close to plain Classical. Since those columns
differ by more than attention alone, the results suggest where a controlled
decoder study would be informative but do not demonstrate that attention causes
the improvement. In particular, the lower collapse rate of attention models
should not be generalized beyond the tested optimizer and implementation.

\subsection{Why the model is not a quantum kernel method}

QEA computes an explicit feature vector for each input. It never evaluates
$k(x_i,x_j)$ between two examples, constructs a Gram matrix, or trains a kernel
classifier. The Pauli vector could be used later to define a projected quantum
kernel, but the attention decoder used here is not such a kernel. Renaming the
method removes an ambiguity that otherwise obscures both novelty and comparison:
the contribution is a controlled audit of a variational quantum embedding, not a
new quantum kernel or a quantum implementation of self-attention.

\subsection{Readout and scaling}
\label{sec:scalability}

The original probability-token design would expose $2^{n_q}$ basis-state
probabilities to dense attention, producing $\mathcal{O}(4^{n_q})$ decoder cost.
The current protocol instead reads all one-body $Z_i$ and two-body $Z_iZ_j$
expectations, so

\begin{equation}
M=n_q+\binom{n_q}{2}=\frac{n_q(n_q+1)}{2},
\qquad
\text{dense-attention cost}=\mathcal{O}(M^2)=\mathcal{O}(n_q^4).
\end{equation}

\begin{table}[htbp]
\centering
\caption{Measured forward cost of the core quantum stage on BCW
(two-repetition ZZ feature map, one-repetition RealAmplitudes ansatz, batch size
as configured). Times are workstation statevector measurements, not hardware
latencies. Circuit depth and readout dimension grow with $n_q$; the observed time
jump at eight qubits reflects this implementation and should not be extrapolated
as a hardware speed model.}
\label{tab:scalability}
\begin{tabular}{rrrrr}
\toprule
$n_q$ & Readout dim $M$ & Circuit depth & Decoder tokens & Fwd/batch (ms) \\
\midrule
2 & 3  & 13 & 3  & 26 \\
4 & 10 & 36 & 10 & 30 \\
6 & 21 & 56 & 21 & 29 \\
8 & 36 & 76 & 36 & 218 \\
\bottomrule
\end{tabular}

\end{table}
\FloatBarrier

Three resources must not be conflated. The classical decoder is polynomial in
$n_q$. Exact statevector simulation still requires $\mathcal{O}(2^{n_q})$ memory
and time up to circuit-depth factors. A quantum device avoids storing that
statevector classically but introduces circuit executions, latency, sampling and
gradient-estimation costs. A short classical output does not by itself rule out a
quantum speedup---many quantum algorithms return few classical values---but this
paper supplies neither a hardness argument for estimating these observables nor
an end-to-end complexity advantage including state preparation and training.

All reported observables commute and can be estimated from the same
computational-basis shots. Each bounded expectation has standard error
$\mathcal{O}(S^{-1/2})$ after $S$ shots; simultaneous control of all $M$
expectations adds a confidence dependence, while parameter-shift training
multiplies circuit evaluations by the number of differentiated parameters. The
paper uses exact expectations and therefore does not measure these costs.

\subsection{Noise, hardware and the early fault-tolerant regime}
\label{sec:hardware}

No hardware or finite-shot result is reported. Gate noise could degrade the
deeper circuits, but it could also interact with optimization as an implicit
regularizer; its direction cannot be inferred from the ideal null result. A
hardware claim would require a specified device, transpiled depth, shot budget,
error-mitigation protocol and repeated comparison against the same classical
surrogate.

Early fault-tolerant devices would relax fidelity and depth constraints, but they
would not automatically validate this architecture. The QEA decoder is classical
and consumes local expectations, so simply running the same shallow circuit with
error correction does not create an algorithmic advantage. Fault-tolerant quantum
linear-algebra transformer proposals use different access assumptions and
block-encoded operations~\cite{zhang2025qtsurvey}; they are not scaled versions of
QEA. A closer continuation would keep more processing on-device and measure class
observables directly, for example
$\ell_c=\langle\psi|O_c|\psi\rangle$, but its trainability, data-loading cost and
resource advantage would still have to be demonstrated rather than assumed.

\subsection{Limitations}
\label{sec:limitations}

The controlled evidence is limited to BCW, five paired seeds and two qubit
counts; the near-ceiling accuracy makes it a weak environment for detecting
component effects. The five-dataset grid is exploratory, contains heterogeneous
readout widths and four shorter-budget rows, and uses QEA-R with a classical
residual bypass. QM9 lacks current-protocol results. BirdCLEF uses a clip-level
rather than recording-grouped split (Appendix~\ref{app:data}), so its estimates
may be optimistic. The BCW and SUSY scalers were also fit before splitting; this
shared, label-free leakage does not favor one paired architecture but can inflate
absolute tabular performance (Appendix~\ref{app:data}). Finally, all quantum results are exact statevector simulations
without finite shots or hardware noise. These limitations prohibit claims of
quantum advantage, hardware readiness, modality-independent superiority, or
equivalence between the PQC and its surrogate.

\section{Conclusion}
\label{sec:conclusion}

Quantum-Embedded Attention provides a clear interface at which to test a PQC
inside a hybrid classifier. The interface-matched BCW factorial does not show a
consistent quantum-minus-classical effect across two circuit widths and two
decoders. One small paired contrast is positive before correction, but it changes
direction at the other width and does not survive correction across the four
embedding contrasts. The experiment is too small and saturated to establish
equivalence. The appropriate conclusion is therefore neither quantum advantage
nor proof of no effect: under this protocol, the PQC contribution is not
replicated across the tested settings.

The exploratory five-dataset grid reinforces the need for that caution. QEA-R is
competitive in point accuracy on three datasets but contains a classical residual
bypass, is worse on SUSY, and fails on CIFAR-10. These observations characterize
the full pipelines; they do not isolate their quantum layers. The most reusable
result is consequently methodological: hybrid models should be evaluated with
paired seeds and controls that preserve the projector, readout interface, decoder
and training budget, and should report failed and incomplete runs explicitly.

The next empirical priority is the same residual-free switch-A comparison on a
non-saturated image or text task, with a prespecified effect size, more seeds,
finite-shot execution and hardware noise. Longer-term, early fault-tolerant work
may study deeper or on-device decoders, but it must include state-preparation,
sampling and training resources in any advantage claim. Those architectures are a
new hypothesis, not an extrapolation of the present results.

\appendix
\section{Backbone and baseline architectures}
\label{app:arch}

This appendix specifies the classical components referenced in
Section~\ref{sec:datasets}. The descriptions follow the configurations used for
the reported rows rather than the defaults of the upstream model families.

\paragraph{Vector backbone (tabular and text).}
BCW and SUSY use an MLP
$\mathbb{R}^{d_\mathrm{in}}\!\to\!128\!\to\!128$ with GELU after each affine
layer, where $d_\mathrm{in}=30$ for BCW and $18$ for SUSY. There is no batch
normalization. AG~News is first encoded by frozen
\texttt{all-MiniLM-L6-v2} into 384-dimensional sentence vectors and then uses
the same $384\!\to\!128\!\to\!128$ MLP. The plain \emph{Classical} baseline
appends a $128\!\to\!C$ linear classifier directly to this backbone, without an
$n_q$ bottleneck.

\paragraph{Vision/audio backbone.}
BirdCLEF spectrograms and CIFAR-10 images use an ImageNet-pretrained ResNeXt-50
(32$\times$4d)~\cite{xie2017resnext}. The configured
\texttt{drop\_last\_blocks: 2} replaces stages 3 and 4 by identity maps; global
average pooling therefore returns a 512-dimensional feature, not the
2048-dimensional output of the complete ResNeXt-50.

\paragraph{Projector and decoder.}
In every Hybrid, Classical + Attention and QEA cell the projector $g_\phi$ is
$d\!\to\!32\!\to\!n_q$ with GELU. The attention decoder uses embedding
dimension 32, four heads, one self-attention block and a 64-unit classifier MLP.
In the factorial, the classical switch-A surrogate is
$\tanh(W\boldsymbol{\xi}+b)$ with $W\in\mathbb{R}^{M\times n_q}$, so the
decoder sees exactly the same $M$-channel interface as for the PQC. It is an
interface control rather than a parameter-count control.

\begin{table}[h]
\centering
\caption{Trainable parameter audit for the BCW factorial. ``Map'' counts the
PQC parameters or the classical switch-A surrogate; ``decoder'' counts the
attention stack or linear classifier. Circuit depth is reported only for the
PQC rows. The small total-count differences are why we use
\emph{interface-matched}, not \emph{parameter-matched}.}
\label{tab:paramcounts}
\begin{tabular}{clrrrr}
\toprule
$n_q$ & Cell & Total & Map & Decoder & Circuit depth \\
\midrule
4 & Quantum + Attention   & 36,014 & 8   & 11,266 & 36 \\
4 & Classical + Attention & 36,056 & 50  & 11,266 & -- \\
4 & Quantum + Linear      & 24,770 & 8   & 22     & 36 \\
4 & Classical + Linear    & 24,812 & 50  & 22     & -- \\
8 & Quantum + Attention   & 36,986 & 16  & 12,098 & 76 \\
8 & Classical + Attention & 37,294 & 324 & 12,098 & -- \\
8 & Quantum + Linear      & 24,962 & 16  & 74     & 76 \\
8 & Classical + Linear    & 25,270 & 324 & 74     & -- \\
\bottomrule
\end{tabular}
\end{table}

\section{Dataset sizes, splits, and leakage caveats}
\label{app:data}

Table~\ref{tab:datastats} records the analysis units represented by the current
result files. Counts are stated separately for train, validation and test so
that caps are not confused with source-dataset sizes.

\begin{table}[h]
\centering
\caption{Dataset statistics for the reported analyses. QM9 is shown for scope
but its available rows use the legacy protocol and are excluded from current
aggregate claims.}
\label{tab:datastats}
\begin{tabular}{lllll}
\toprule
Dataset & Input to backbone & Train/val/test used & Classes & Split construction \\
\midrule
BCW      & 30 features & 397/86/86 & 2 & stratified 70/15/15 \\
SUSY     & 18 features & 14,000/3,000/3,000 & 2 & stratified 70/15/15 \\
AG~News  & MiniLM 384 & 20,000/2,000/2,000 & 4 & official train/test \\
CIFAR-10 & $224{\times}224$ RGB & 20,000/4,000/4,000 & 10 & official train/test \\
BirdCLEF & 5-s mel spectrogram & 1,920/240/240 & 2 & clip-level fold \\
QM9      & 14 descriptors & legacy rows only & 5 bins & excluded here \\
\bottomrule
\end{tabular}
\end{table}

For AG~News, 20,000 examples are selected from the official training split; the
first 4,000 examples of the official test split are divided equally and
stratified into validation and test. For CIFAR-10, 15\% of the official 50,000
training images form the validation pool before caps are applied; the retained
sets contain 20,000 training, 4,000 validation and 4,000 official-test images.
The SUSY loader reads the first 20,000 finite rows of the local UCI file.

\paragraph{Feature-scaling leakage.}
The BCW and SUSY loaders used for the archived experiments fit
\texttt{StandardScaler} to the complete sampled feature matrix before the
train/validation/test split. This does not use labels and is identical across
paired model cells, so it does not explain a within-seed quantum--classical
difference. It nevertheless transfers validation/test feature moments into
training and can make absolute generalization estimates optimistic. The reported
tabular scores must therefore be read as internal comparisons under this shared
preprocessing pipeline, not as leakage-free benchmark estimates. A definitive
rerun should split first, fit the scaler on training data only, and transform
validation and test with those fixed statistics.

\paragraph{BirdCLEF split (recording leakage caveat).}
The BirdCLEF task is built by cutting soundscape recordings into five-second
clips. The train/evaluation split is stratified over clips (using the
\texttt{nocall} label), not grouped by source recording. Clips from the same
recording can therefore occur on both sides of the split, which may bias every
BirdCLEF estimate upward. A recording-grouped split is required for a definitive
audio comparison; the present BirdCLEF row is treated as exploratory.

\bmhead{Reproducibility statement}
The source archive, configuration grid (including the interface-matched
factorial), run-level CSV files, and analysis scripts that regenerate the reported
tables and figures accompany the revision. Random seeds and dependency versions
are recorded in the configurations and lockfile.

\bmhead{Data availability}
All six source datasets are publicly available from the repositories cited in
Section~\ref{sec:datasets}. The run-level measurements supporting the five-dataset
cross-modality analysis and the BCW factorial accompany the revision as CSV files.

\bmhead{Code availability}
The implementation and exact experiment configurations are available at
\url{https://github.com/MarkCodering/quantum-embedding-with-transformer} and are
also supplied with the revision so that review does not depend on repository
availability.

\bmhead{Competing interests}
This declaration must be completed and approved by all authors before
resubmission.

\bmhead{Acknowledgments}
Omitted for anonymous review.

\bibliography{references}

@article{preskill2018nisq,
  author  = {Preskill, John},
  title   = {Quantum Computing in the {NISQ} era and beyond},
  journal = {Quantum},
  year    = {2018},
  volume  = {2},
  pages   = {79},
  doi     = {10.22331/q-2018-08-06-79}
}

@article{benedetti2019pqc,
  author  = {Benedetti, Marcello and Lloyd, Erika and Sack, Stefan and Fiorentini, Mattia},
  title   = {Parameterized quantum circuits as machine learning models},
  journal = {Quantum Science and Technology},
  year    = {2019},
  volume  = {4},
  number  = {4},
  pages   = {043001}
}

@article{mcclean2018bp,
  author  = {McClean, Jarrod R. and Boixo, Sergio and Smelyanskiy, Vadim N. and Babbush, Ryan and Neven, Hartmut},
  title   = {Barren plateaus in quantum neural network training landscapes},
  journal = {Nature Communications},
  year    = {2018},
  volume  = {9},
  pages   = {4812}
}

@article{schuld2020vqc,
  author  = {Schuld, Maria and Bocharov, Alex and Svore, Krysta M. and Wiebe, Nathan},
  title   = {Circuit-centric quantum classifiers},
  journal = {Physical Review A},
  year    = {2020},
  volume  = {101},
  pages   = {032308}
}

@article{havlicek2019supervised,
  author  = {Havl{\'\i}{\v{c}}ek, Vojt{\v{e}}ch and C{\'o}rcoles, Antonio D. and Temme, Kristan and Harrow, Aram W. and Kandala, Abhinav and Chow, Jerry M. and Gambetta, Jay M.},
  title   = {Supervised learning with quantum-enhanced feature spaces},
  journal = {Nature},
  year    = {2019},
  volume  = {567},
  pages   = {209--212}
}

@article{schuld2021kernel,
  author  = {Schuld, Maria and Killoran, Nathan},
  title   = {Quantum machine learning in feature {Hilbert} spaces},
  journal = {Physical Review Letters},
  year    = {2019},
  volume  = {122},
  pages   = {040504}
}

@article{perez2020data,
  author  = {P{\'e}rez-Salinas, Adri{\'a}n and Cervera-Lierta, Alba and Gil-Fuster, Elies and Latorre, Jos{\'e} I.},
  title   = {Data re-uploading for a universal quantum classifier},
  journal = {Quantum},
  year    = {2020},
  volume  = {4},
  pages   = {226}
}

@article{henderson2020quanvolutional,
  author  = {Henderson, Maxwell and Shakya, Samriddhi and Pradhan, Shashindra and Cook, Tristan},
  title   = {Quanvolutional neural networks: powering image recognition with quantum circuits},
  journal = {Quantum Machine Intelligence},
  year    = {2020},
  volume  = {2},
  pages   = {2}
}

@article{cherrat2024qsa,
  author  = {Cherrat, El Amine and Kerenidis, Iordanis and Mathur, Natansh and Landman, Jonas and Strahm, Martin Felix and Li, Yun Yvonna},
  title   = {Quantum vision transformers},
  journal = {Quantum},
  year    = {2024},
  volume  = {8},
  pages   = {1265}
}

@misc{qiskitml,
  author = {{Qiskit Machine Learning Developers}},
  title  = {Qiskit Machine Learning},
  year   = {2024},
  url    = {https://qiskit-community.github.io/qiskit-machine-learning/}
}

@article{bergholm2018pennylane,
  author  = {Bergholm, Ville and others},
  title   = {{PennyLane}: Automatic differentiation of hybrid quantum-classical computations},
  journal = {arXiv:1811.04968},
  year    = {2018}
}

@article{chen2024qet,
  author  = {Chen, Hao-Yuan and Chang, Yen-Jui and Liao, Shih-Wei and Chang, Ching-Ray},
  title   = {Quantum Embedding with Transformer for High-dimensional Data},
  journal = {arXiv:2402.12704},
  year    = {2024}
}

@techreport{krizhevsky2009cifar,
  author      = {Krizhevsky, Alex},
  title       = {Learning Multiple Layers of Features from Tiny Images},
  institution = {University of Toronto},
  year        = {2009}
}

@article{wolberg1995bcw,
  author  = {Street, W. Nick and Wolberg, William H. and Mangasarian, Olvi L.},
  title   = {Nuclear feature extraction for breast tumor diagnosis},
  journal = {SPIE Biomedical Image Processing and Biomedical Visualization},
  year    = {1993},
  volume  = {1905},
  pages   = {861--870}
}

@article{ramakrishnan2014qm9,
  author  = {Ramakrishnan, Raghunathan and Dral, Pavlo O. and Rupp, Matthias and von Lilienfeld, O. Anatole},
  title   = {Quantum chemistry structures and properties of 134 kilo molecules},
  journal = {Scientific Data},
  year    = {2014},
  volume  = {1},
  pages   = {140022}
}

@article{baldi2014susy,
  author  = {Baldi, Pierre and Sadowski, Peter and Whiteson, Daniel},
  title   = {Searching for exotic particles in high-energy physics with deep learning},
  journal = {Nature Communications},
  year    = {2014},
  volume  = {5},
  pages   = {4308}
}

@inproceedings{zhang2015agnews,
  author    = {Zhang, Xiang and Zhao, Junbo and LeCun, Yann},
  title     = {Character-level convolutional networks for text classification},
  booktitle = {Advances in Neural Information Processing Systems (NeurIPS)},
  year      = {2015},
  volume    = {28}
}

@inproceedings{wang2020minilm,
  author    = {Wang, Wenhui and Wei, Furu and Dong, Li and Bao, Hangbo and Yang, Nan and Zhou, Ming},
  title     = {{MiniLM}: Deep self-attention distillation for task-agnostic compression of pre-trained transformers},
  booktitle = {Advances in Neural Information Processing Systems (NeurIPS)},
  year      = {2020},
  volume    = {33}
}

@inproceedings{xie2017resnext,
  author    = {Xie, Saining and Girshick, Ross and Doll{\'a}r, Piotr and Tu, Zhuowen and He, Kaiming},
  title     = {Aggregated residual transformations for deep neural networks},
  booktitle = {IEEE Conference on Computer Vision and Pattern Recognition (CVPR)},
  year      = {2017},
  pages     = {1492--1500}
}

@inproceedings{kingma2015adam,
  author    = {Kingma, Diederik P. and Ba, Jimmy},
  title     = {{Adam}: A method for stochastic optimization},
  booktitle = {International Conference on Learning Representations (ICLR)},
  year      = {2015}
}

@article{birdclef2021,
  author  = {Kahl, Stefan and Denton, Tom and Klinck, Holger and Glotin, Herv{\'e} and Go{\"e}au, Herv{\'e} and Vellinga, Willem-Pier and Planqu{\'e}, Robert and Joly, Alexis},
  title   = {Overview of {BirdCLEF} 2021: Bird call identification in soundscape recordings},
  journal = {CLEF Working Notes},
  year    = {2021}
}

@article{schuld2021effect,
  author  = {Schuld, Maria and Sweke, Ryan and Meyer, Johannes Jakob},
  title   = {Effect of data encoding on the expressive power of variational quantum-machine-learning models},
  journal = {Physical Review A},
  year    = {2021},
  volume  = {103},
  pages   = {032430}
}

@article{schnabel2025scrutiny,
  author  = {Schnabel, Jan and Roth, Marco},
  title   = {Quantum kernel methods under scrutiny: a benchmarking study},
  journal = {Quantum Machine Intelligence},
  year    = {2025},
  volume  = {7},
  number  = {1},
  pages   = {58},
  doi     = {10.1007/s42484-025-00273-5}
}

@article{zhang2025qtsurvey,
  author  = {Zhang, Hui and Zhao, Qinglin},
  title   = {A Survey of Quantum Transformers: Approaches, Advantages, Challenges, and Future Directions},
  journal = {arXiv:2504.03192},
  year    = {2025}
}

\end{document}